\documentclass[twocolumn,showpacs,preprintnumbers,amsmath,amssymb]{revtex4}

\usepackage{graphicx}
\usepackage{dcolumn}
\usepackage{bm}

\begin{document}

\title{Minimal timing jitter in superconducting nanowire single photon detectors}

\author{D. Yu. Vodolazov}

\email{vodolazov@ipmras.ru}

\affiliation{Institute for Physics of Microstructures, Russian
Academy of Sciences, 603950, Nizhny Novgorod, GSP-105, Russia}

\affiliation{Department of Physics and IT, Moscow Pedagogical State University, 119991 Moscow, Russia}

\date{\today}

\begin{abstract}

Using two-temperature model coupled with modified time-dependent
Ginzburg-Landau equation we calculate the delay time $\tau_d$ in
appearance of growing normal domain in the current-biased
superconducting strip after absorption of the single photon. We
demonstrate that $\tau_d$ depends on the place in the strip where
photon is absorbed and monotonically decreases with increasing of
the current. We argue, that the variation of $\tau_d$ (timing
jitter), connected either with position-dependent response or Fano
fluctuations could be as small as the lowest relaxation time of
the superconducting order parameter $\sim \hbar/k_BT_c$ ($T_c$ is
the critical temperature of the superconductor) when the current
approaches the depairing current.
\end{abstract}

\maketitle

\section{Introduction}

In superconducting nanowire single photon detector (SNSPD)
absorption of single photon of visible or infra-red range with
energy $E_{\nu}$ leads to appearance of voltage pulse at
relatively large transport current in superconducting strip.
Experiments demonstrate that there is a finite delay time $\tau_d$
in appearance of the voltage response and moreover there is a
variance in $\tau_d$ (called as a timing jitter) which depends on
the material or bias current
\cite{Zhang,Natarajan,Calandri,Sidorova,Caloz,Korzh}. The origin
for the timing jitter may come from the electronics, read-out
system or finite length of the strip (geometrical jitter
\cite{Calandri}) but it also may have an intrinsic origin
connected with dynamics of the superconducting order parameter
$\Delta$ in response of current-carrying superconducting strip on
absorbed photon. Indeed, the photon heats electrons (theoretical
estimations show that locally the electron temperature may well
exceed critical temperature of the superconductor
\cite{Vodolazov_PRA,Kozorezov_Fano}) but due to finite relaxation
time of magnitude of $\Delta$ ($\tau_{|\Delta|}$) the
superconductivity is not destroyed instantly. This effect is
well-known, for example, from the study of time response of
superconducting bridge/stripe on the supercritical current pulse
(current pulse with an amplitude exceeding critical current)
\cite{Tinkham_time_delay,Pals,Geier,Jelila,Vodolazov_time_delay}.
In that works it was found finite delay time which is strongly
reduced with increasing of the current pulse amplitude -
qualitatively similar result is found in experiments with SNSPD
\cite{Zhang,Sidorova,Caloz,Korzh}.

In SNSPD timing jitter could be connected with position-dependent
response \cite{Vodolazov_PRA,Zotova_SUST,Engel_IEEE_model}, when
photon absorbed in different sites across the strip produces the
voltage signal at the different detection (critical) currents
$I>I_{det}(y)$ ($y$ is a coordinate across the strip). Then in
accordance with results of
Refs.\cite{Tinkham_time_delay,Pals,Geier,Jelila,Vodolazov_time_delay}
one may expect the different delay time at fixed current $I$:
$\tau_d(I/I_{det}(y))$, depending where in the superconductor the
photon is absorbed.

Additional mechanism of timing jitter in SNSPD comes from
so-called Fano-fluctuations \cite{Kozorezov_Fano} (lose of the
part of energy of the photon due to fluctuated nature of escape of
nonequilibrium Debye phonons to the substrate) or local variations
of material parameters of the superconducting strip (mean free
path, local $T_c$ or thickness of the strip). Because local
detection current $I_{det}(y)$ depends on the deposited energy $E$
to the electron/phonon system (it determines how strong electrons
and phonons are heated) and on the material parameters, at fixed
current the ratio $I/I_{det}$ varies from one absorption event to
another one and it produces the variance in the delay time.

In this paper, based on the two-temperature model coupled with
modified time-dependent Ginzburg-Landau equation and current
continuity equation \cite{Vodolazov_PRA} we calculate the
position-dependent delay time in SNSPD both in absence and in
presence of Fano fluctuations and study how $\tau_d$ depends on
the current and deposited energy. Effect of Fano fluctuations in
our model are taken into account via introduction of probability
$P(E)$ to deposit energy $E<E_{\nu}$ to the electron/phonon
systems of superconductor \cite{Kozorezov_Fano,Kozorezov_arxiv}.
Effect of variations of material parameters may be considered in a
similar way \cite{Kozorezov_arxiv} and we do not study them
explicitly. We define the delay time $\tau_d$ as a time needed for
formation of the growing normal domain after absorption of the
photon somewhere in the superconducting strip. We find, that
$\tau_d$ is drastically reduced as the current approaches to the
depairing current and timing jitter may be as small as
$\hbar/k_BT_c$ ($\sim $ 0.8 ps for superconductor with $T_c = 10
K$). We also show that the considered model with
position-dependent response predicts stronger deviation of
dependence of photon counts on the delay time (in the literature
its is called as probability density function (PDF)
\cite{Sidorova} or instrument response function
(IRF)\cite{Korzh,Kozorezov_arxiv}) from the Gaussian-like
distribution than the hot belt model predicts
\cite{Kozorezov_arxiv}. We argue that it occurs due to photons
absorbed near the edge of the strip which provide the largest
delay time.

The structure of the paper is following. In section II we introduce our theoretical model.
In section III we present our results on dependence of position-dependent $\tau_d$ on current
in absence of Fano fluctuations and in section IV we include effect
of Fano fluctuations and calculate energy dependence of $\tau_d$.
In section V we discuss the value of delay time and timing jitter at low currents, when intrinsic
detection efficiency of the detector is much smaller than unity and in section VI we relate our results
with existing experiments and theoretical works.

\section{Model}

The main assumption of our model implies that in any moment of
time distribution function of electrons and phonons could be
described by Fermi-Dirac and Bose-Einstein functions with local
temperatures of electrons $T_e$ and phonons $T_p$ different from
the bath temperature $T$. In Ref. \cite{Vodolazov_PRA} it is shown
that this assumption is approximately valid in rather dirty (with
diffusion coefficient $D \lesssim 0.5 cm^2/s$) thin
superconducting films and energy of photon $E_{\nu} \gtrsim 1eV$.
In this model temporal and space evolution of $T_e$ and $T_p$ is
governed by heat conductance and energy balance equations (Eq.
(30) and (31), respectively, in \cite{Vodolazov_PRA}) where heat
capacity and heat conductivity (Eq. (31) in \cite{Vodolazov_PRA})
of electrons take into account the presence of the superconducting
gap. These equations are coupled to the time-dependent
Ginzburg-Landau (TDGL) equation for superconducting order
parameter $\Delta$ (Eq. (36) in \cite{Vodolazov_PRA}) which is
modified to take into account correct temperature dependence of
coherence length, superconducting order parameter and
critical(depairing) current at temperatures far below $T_c$.
Together with these equations one also has to solve current
continuity equation - Eqs. (37) in Ref. \cite{Vodolazov_PRA}.

In this model the absorbed photon is modelled by instant local
heating of both electrons and phonons up to $T_e=T_p=T_{eff}$ in
the area $2\xi_c\times 2\xi_c$ - so called initial hot spot
\cite{Vodolazov_PRA}, where $T_{eff}$ should be determined from
the energy conservation law (Eq. (15) in \cite{Vodolazov_PRA}).
Here $\xi_c=(\hbar D/k_BT_c)^{1/2} \sim \xi_0=(\hbar
D/1.76k_BT_c)^{1/2}$ is convenient in numerical calculations
length scale, the initial hot spot is placed at $x=L/2$ and
different transverse coordinates $y=0 - w$ ($L$ is a length of the
strip and $w$ is its width). In Ref. \cite{Vodolazov_PRA} it is
discussed the eligibility and limitation of this choice of initial
condition on the basis of kinetic equations approach.

In numerical calculations we use parameters of typical NbN strip:
$w=20 \xi_c\simeq 130 nm$ ($\xi_c=6.4 nm$), thickness $d=4 nm$,
$T_c=10 K$. Important parameters $\gamma=10$ and $\tau_0=900 ns$
which stay in front of electron-phonon and phonon-electron
collision integrals in kinetic equations (see Eqs. (3,4,6,7) and
Eqs (30,31) in \cite{Vodolazov_PRA}) control corresponding
electron-phonon $\tau_{ep}$ and phonon-electron $\tau_{pe}$
inelastic relaxation times. We also use $L=4w=80 \xi_c$,
$\tau_{esc}=0.05 \tau_0$ (escape time of nonequilibrium phonons to
the substrate) and the boundary conditions for $\Delta$, $T_e$ and
electrostatic potential $\varphi$ in $x$ and $y$ directions from
Ref. \cite{Vodolazov_PRA}.

Strictly speaking TDGL equation was derived at temperatures close
to $T_c$, and it is quantitatively valid when $|\Delta|<k_BT_e$
\cite{Watts-Tobin,Schmid}. Note that in the hot spot area
the local temperature satisfies this condition (at least at the initial
stage of its time evolution) and to the moment of appearance of
first vortices does not strongly deviate from $T_c$ (see for
example Fig. 8 in \cite{Vodolazov_PRA}). Secondly, we also did our
calculations at temperatures close to $T_c$ (at $T/T_c=$0.9 and
0.95) and did not find any qualitative difference with
results found at lower $T$. Namely, when initial hot spot appears
in the central part of the strip the vortex-antivortex nucleate in
that place and move in opposite directions, while when it
appears near the edge the vortex enters the strip when $I>I_{det}(y)$.
The only quantitative difference is that
at $T \sim T_c$ delay time becomes much larger (which favors
energy leakage of photon's energy to substrate) and due to lower
absolute value of detection current the normal domain
grows much slowly or even does not appear in the strip (depending
on choice of $\tau_{esc}$ and $T$). As a result the order
parameter relaxes in hot spot area to the equilibrium value after
passage of several vortices(antivortices) across the strip without
appearance of large voltage signal.
\begin{figure}[hbtp]
\includegraphics[width=0.55\textwidth]{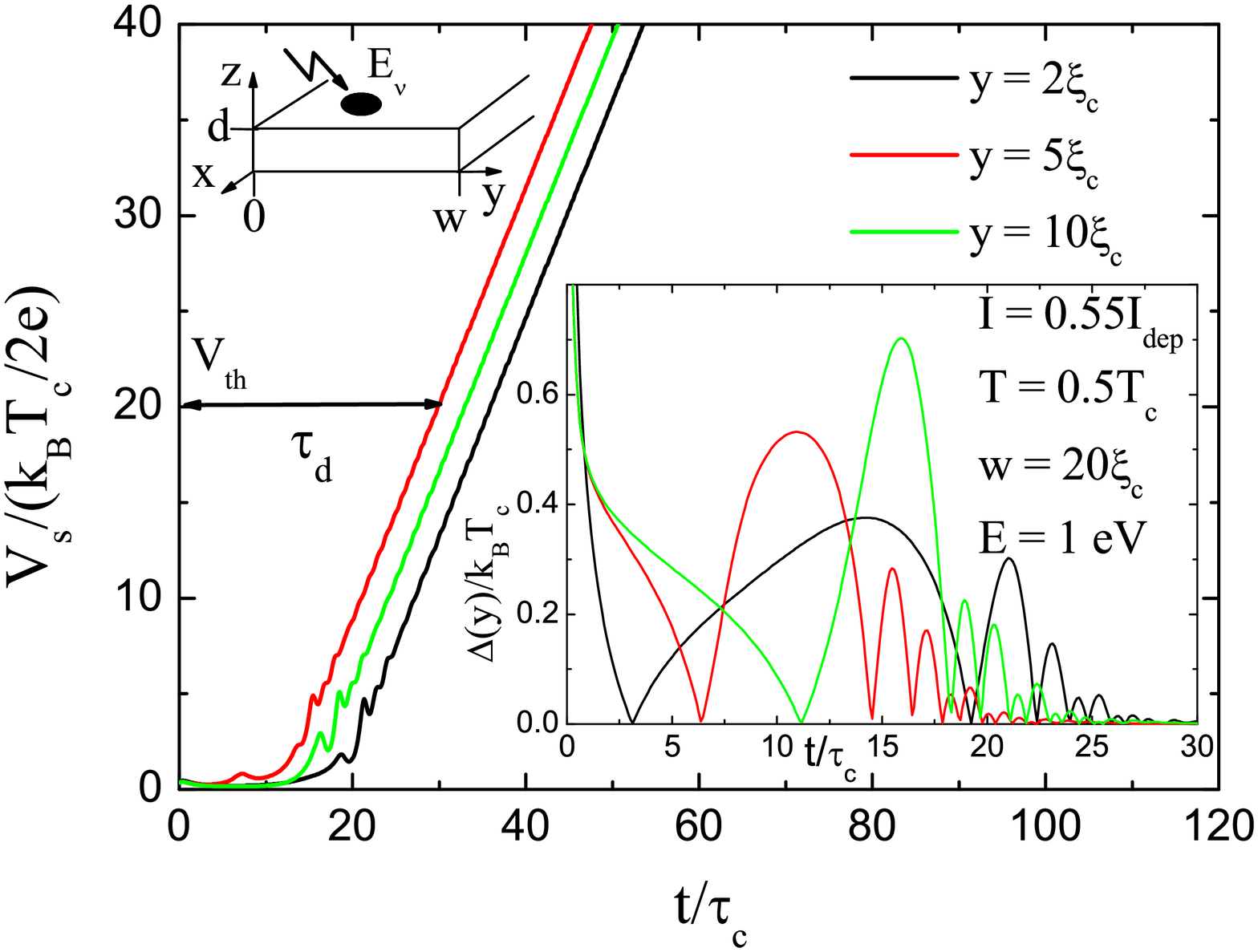}
\caption{Time dependence of the voltage drop along the superconducting
strip (time is normalized in units of $\tau_c=\hbar/k_BT_c$).
The initial hot spot appears at $t=0$ in different places
($y= 2,5,10 \xi_c$) across the strip with width
20$\xi_c$. The definition of $\tau_d$ is seen in the figure. In
the right inset we show time dependence of $\Delta$ in the center of
initial hot spot. Deposited energy to electron/phonon systems
$E = 1 eV$ corresponds to
initial temperature $T_{init}=2T_c$ (see section II).
In the left inset we show the geometry of the strip.}
\end{figure}

In our work we do not consider fluctuation assisted photon
counting at $I<I_{det}$ which is connected with penetration vortices via the energy barrier formed
near the hot spot (so we are working strictly in
so-called deterministic regime \cite{Sidorova}).
Therefore delay time is not defined at $I<I_{det}(y)$
and it is finite at $I \geq I_{det}(y)$ (see section III).

\section{Position-dependent delay time}

In Fig. 1 we show time dependence of the voltage response of superconducting strip
after appearance in the superconductor of the initial hot spot at $t=0$.
One can see that depending on the site of the initial hot spot (associated with the
site where the photon is absorbed) there is different delay time in
appearance of large (growing linearly in
time) voltage, connected with appearance of the growing
normal domain. From these dependencies it is clear that the variance
in $\tau_d$ does not depend on choice of threshold voltage $V_{th}$ (if it is large enough)
and in the following we choose level $V_{th}=20 V_c=20 k_BT_c/2e$ for
quantitative evaluation of $\tau_d$.
\begin{figure}[hbtp]
\includegraphics[width=0.5\textwidth]{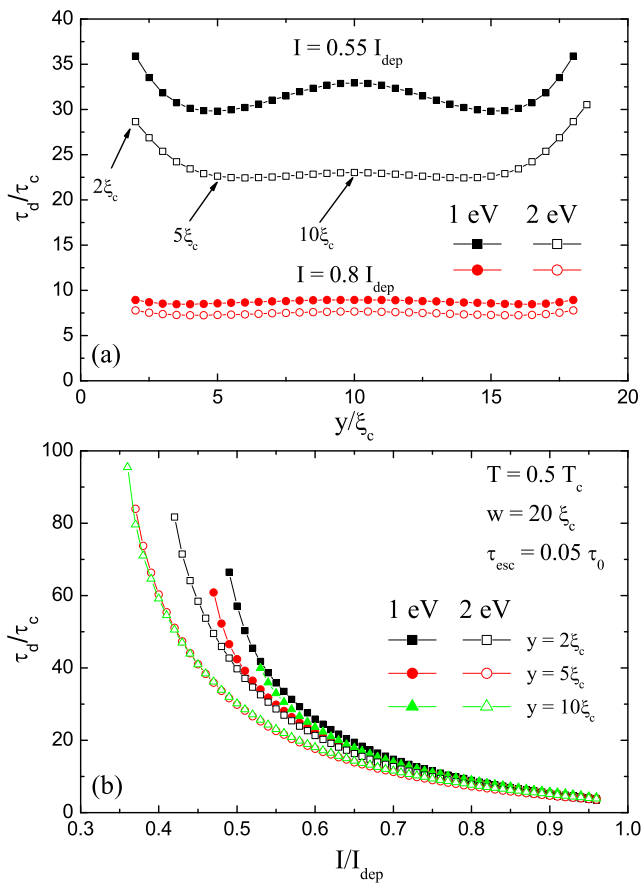}
\caption{(a) Position-dependent $\tau_d$ at different currents and
deposited energies ($E = 1 eV \to T_{init}=2T_c $ and
$E = 2 eV \to T_{init}=2.4T_c$). (b) Dependence of $\tau_d$
on current for three positions of initial hot spot $y=2, 5, 10
\xi_c$ and two deposited energies 1 and 2 eV.}
\end{figure}

In Fig. 2(a,b) we present position and current dependence of
$\tau_d$. Delay time is minimal in the place where $I_{det}(y)$ is
minimal (compare Fig. 2a with Fig. 3a) and $\tau_d$ monotonically
decreases with current increase (at fixed position of hot spot -
see Fig. 2b). Both these results are not surprising and resemble
previous theoretical findings on the time delay in destruction of
superconducting state by current pulse
\cite{Tinkham_time_delay,Pals,Geier,Vodolazov_time_delay}.
According to these results $\tau_d$ monotonically decreases with
increase of ratio $I/I_c$, where $I_c$ is the critical current of
the superconducting bridge/strip. In our problem role of $I_c$ is
played by $I_{det}(y)$ and situation is more complicated, because
we are looking not for the suppression of superconductivity
(appearance of the phase slip center/line or moving vortices) but
for nucleation of the growing normal domain. From Fig. 1 it is
clear that these are not the same. For example at $y=2\xi_c$
(photon is absorbed near the edge of the strip) the vortex appears
earlier than the vortex/antivortex pair nucleates at $y=10 \xi_c$
(photon is absorbed in the center of the strip) because $I_{det}$
in that place smaller but the normal domain appears earlier in the
last case due to shorter time needed to cross the strip by the
vortex and antivortex than by single vortex (only after that the
normal domain appears and expands along the strip, leading to
large voltage response). Moreover in the considered model the
appearance of the vortices does not obligatory lead to appearance
of the normal domain when the bias current is close to the
retrapping current (see discussion in Ref. \cite{Vodolazov_PRA}).
Therefore in our 2D case $\tau_d$ is not only function of ratio
$I/I_{det}(y)$, but it may also depend on location of initial hot
spot. For example at $y=2 \xi_c$ detection current is the same as
at $y \sim 7 \xi_c$ (see Fig. 3a in case of $E = 1 eV$) but
$\tau_d$ are different (see Fig. 2a).
\begin{figure}[hbtp]
\includegraphics[width=0.5\textwidth]{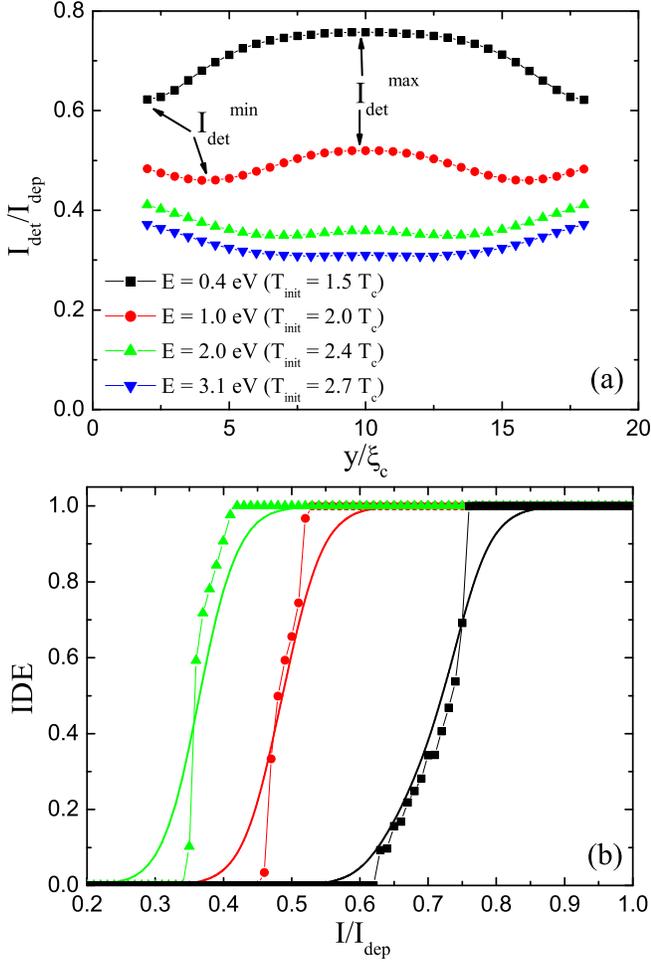}
\caption{(a) Position-dependent detection current for different
deposited energies $E$. (b) Dependence of intrinsic detection
efficiency (IDE) on the current for three values of deposited
energy 0.4, 1 and 2 eV. Symbols are obtained using results shown
in (a) assuming equal probability of photon absorption across the
width of the strip and no fluctuations in the deposited energy.
Solid curves are obtained in presence of both position-dependent
$I_{det}(y)$ and Fano fluctuations which provide local
fluctuations of $I_{det}(y)$. In our model instead of
error-function \cite{Kozorezov_Fano} we use simpler expression for
local detection efficiency $LDE(y)=0.5\cdot(1+{\rm
tanh}((I-I_{det}(y))/dI))$ with control parameter $dI=0.05
I_{dep}$ ($IDE(I)=\int_0^w LDE(y)dy/w$) to show qualitatively
effect of Fano fluctuations. When Fano fluctuations are absent
($dI=0$) one comes to curves with symbols.}
\end{figure}

In contrast to problem with supercritical current pulse
\cite{Tinkham_time_delay,Pals,Geier,Vodolazov_time_delay} $\tau_d$
does not diverge as $I\to I_{det}(y)$ (see Fig. 2b). This fact is
connected with dynamic nature of the hot spot and its finite
life-time. When the hot spot expands the electronic temperatures
goes down, but size of hot spot increases and there is an
'optimal' hot spot (with 'optimal' size and 'optimal' $T_e$) for
given deposited energy $E$ which provides the 'minimal' detection
current (do not confuse it with $I_{det}^{min}$ in Fig. 3a) for
fixed photon absorption site. But there is also finite relaxation
time of $|\Delta|$, leading to finite $\tau_d$ which grows with
decreasing of the current. So, roughly speaking the maximal
$\tau_d$ in Fig. 2b (and corresponding current is the 'minimal'
detection current those coordinate dependence is shown in Fig. 3a)
corresponds to the time needed for the growing hot spot to become
the 'optimal' one.
\begin{figure}[hbtp]
\includegraphics[width=0.5\textwidth]{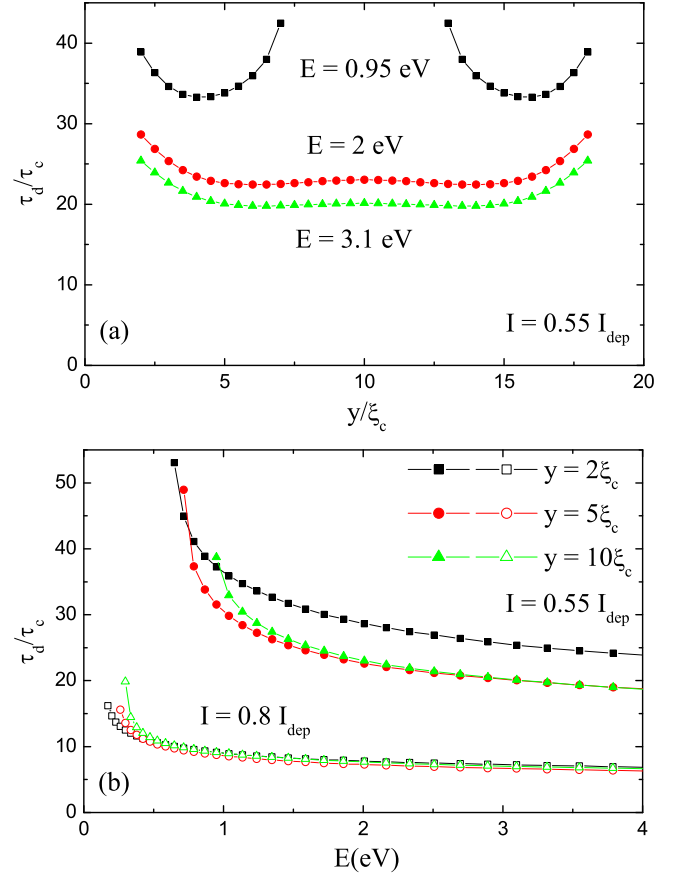}
\caption{(a) Position-dependent $\tau_d$ for different deposited
energies $E$ and $I=0.55 I_{dep}$. At $E = 0.95 eV$ central part of the strip
does not 'detect' photons (absorbed photon does not 'produce' vortices and
normal domain does not appear) and formally $\tau_d=\infty$. (b) Dependence
of $\tau_d$ on deposited energy for three photon's absorption cites $y=2,5,10
\xi_c$ and two values of the current.}
\end{figure}

From Fig. 2a,b one can see that with increasing of the current the variance in delay time
decreases and it approaches $\hbar/k_BT_c$ as the current goes to the depairing current.
In the next section we discuss how Fano fluctuations affect this result.

\section{Delay time in presence of Fano fluctuations}

In this section we consider effect of Fano fluctuations on
delay time and timing jitter. We follow the Ref. \cite{Kozorezov_arxiv}
and introduce normalized probability of energy deposition $E$ both
to electron and phonon systems after absorption of the photon
\begin{equation}
P(E)=\frac{1}{\sigma\sqrt{2\pi}}e^{-(E-\bar{E})^2/2\sigma^2}.
\end{equation}

In this model it is assumed that the part of energy of photon
$E_{\nu}-E$ is lost due to fluctuations in escape of
nonequilibrium Debye phonons to the substrate
\cite{Kozorezov_Fano,Kozorezov_arxiv} and the most probable
deposited energy is equal to $E={\bar E}<E_{\nu}$. In Fig. 4a we
show position-dependent delay time for different $E$ and in Fig 4b
we demonstrate energy dependence of $\tau_d$ at different sites
where photon is absorbed. Based on these results and Eq. (1) we
calculate and plot in Fig. 5 the local probability to observe some
$\tau_d$ in case of absorption of the photon in the center of the
strip $P(\tau_d,y=10 \xi_c)$ (for this purpose we convert
dependence $\tau_d(E)$ to $E(\tau_d)$ and insert it to Eq. (1)).
In calculations we use $\bar{E}=1.5, 2.5 eV$ and $\sigma=0.1
\bar{E}$. One can see that with increasing of the current (at
fixed $\bar{E}$) or $\bar{E}$ (at fixed current) the function
$P(\tau_d,y)$ tends to Gaussian-like form. This result follows
from the dependence $\tau_d(E)$ - at large currents and $E$ it is
better approximated by linear dependence $\tau_d(E)=a+bE$ in the
finite range of energies $\sim 2\sigma$ which together with Eq.
(1) automatically leads to Gaussian-like dependence. Because
nonlinearity is stronger at smaller $E$ (large $\tau_d$)
dependence $P(\tau_d,y)$ is not Gaussian-like at large $\tau_d$.

\begin{figure}[hbtp]
\includegraphics[width=0.5\textwidth]{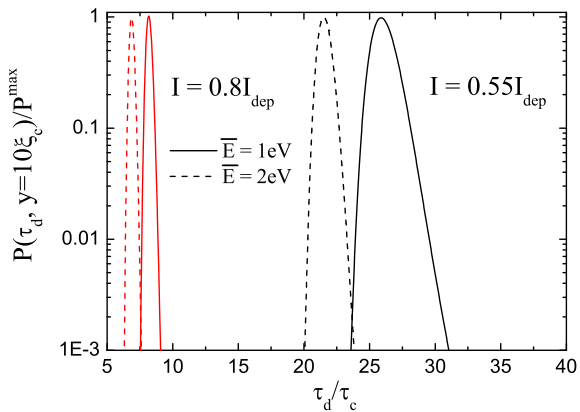}
\caption{Local normalized probability to have delay time $\tau_d$ at different currents
and different ${\bar E}\sim E_{\nu}$ (solid curves - ${\bar E}=1.5 eV$, dashed curves -
${\bar E}=2.5 eV$) for the photon absorbed in the center of the strip ($y = 10 \xi_c$).
In calculations we take $\sigma=0.1 {\bar E}$.}
\end{figure}

Now we can combine this result with position-dependent $\tau_d$.
We calculate $P(\tau_d,y)$ at each discrete point of our numerical
grid, integrate it over the $y$ and assume equal probability for
photon absorption across the strip. In this way we find
$P(\tau_d)=\int P(\tau_d,y)dy$ which is proportional to
experimentally found probability density function \cite{Sidorova},
instrument response function \cite{Korzh,Kozorezov_arxiv} or
dependence of photon counts on delay time - see Fig. 6. Local
$P(\tau_d,y)$ at any $y$ has the shape similar to the one shown in
Fig. 5 but centered at different $\tau_d$. Contribution from the
near edge regions, which give the largest $\tau_d$, provides on
dependence $P(\tau_d)$ some kind of 'shoulder' at relatively small
current $I=0.55 I_{dep}$ ('oscillations' on the 'shoulder' visible
for ${\bar E}=2.5 eV$ have artificial origin and are connected
with finite step $dy=0.5 \xi_c$ used in numerical calculations),
while at relatively large current the 'shoulder' practically
disappears. Visibility of the 'shoulder' depends on parameter
$\sigma$ in Eq. 1 and with its increase the 'shoulder' smears out,
leading to shape of $P(\tau_d)$ qualitatively similar to the one
shown in Fig. 5, while with its decrease the 'shoulder' becomes
more pronounced.
\begin{figure}[hbtp]
\includegraphics[width=0.5\textwidth]{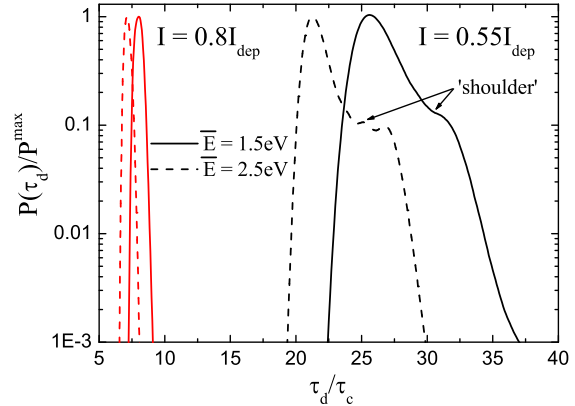}
\caption{Normalized probability to have delay time $\tau_d$
(parameters are the same as in Fig. 5). At relatively low current
($I=0.55 I_{dep}$) there is a 'shoulder' on dependence $P(\tau_d)$
connected with contribution of photons absorbed in near-edge regions
of the strip which provide large $\tau_d$.}
\end{figure}

From Fig. 4b and 6 it follows that timing jitter in presence of both position
dependent-response and relatively large Fano fluctuations ($\sigma = 0.1 {\bar E}$)
 still could be about of $\hbar/k_BT_c$
(when deposited energy ${\bar E} > 1eV$) as the current approaches
to the depairing current. Physically it is connected with relatively
short delay time when $I/I_{det} \gtrsim 1.8 $ (see section Discussion
below) which is the case for our parameters (see Fig. 3a)
as $I\to I_{dep}$.

\section{Jitter at low detection efficiency}

So far we consider delay time and timing jitter at currents
exceeding $I_{det}^{max}$ (see Fig. 3a) when intrinsic detection
efficiency reaches unity (or photon count rate (PCR), system
detection efficiency (SDE) reaches the plateau or saturate at
relatively large current). At $I>I_{det}^{max}$ both $\tau_d$ and
timing jitter decreases with increasing of the current. What one
can expect at low currents $I\gtrsim I_{det}^{min}$ when IDE $\ll
1$?

In the model with position dependent response and no Fano
fluctuation the detector stops to operate at $I<I_{det}^{min}$. At
current slightly exceeding $I_{det}^{min}$ only part of the strip
where $I>I_{det}(y)$ detects photons and it is clear that position
dependent timing jitter in this case should be small. To
illustrate it in Fig. 7 we show $\tau_d$ at different currents
just above $I_{det}^{min}$.

In presence of Fano fluctuations $I_{det}^{min}$ varies from one
act of photon's absorption to another one because of variation of
the deposited energy $E$. But the maximal deposited energy cannot
exceed the energy of the photon $E_{\nu}$ and, hence, there is a
minimal $I_{det}^{min}$ which corresponds to $E=E_{\nu}$. The same
situation is with variation of material parameters - in the
'weakest' place of the strip $I_{det}$ reaches the minimal value
when $E=E_{\nu}$. Therefore in framework of the used model we
expect that at low currents timing jitter decreases (while delay
time increases) with decrease of the current.
\begin{figure}[hbtp]
\includegraphics[width=0.55\textwidth]{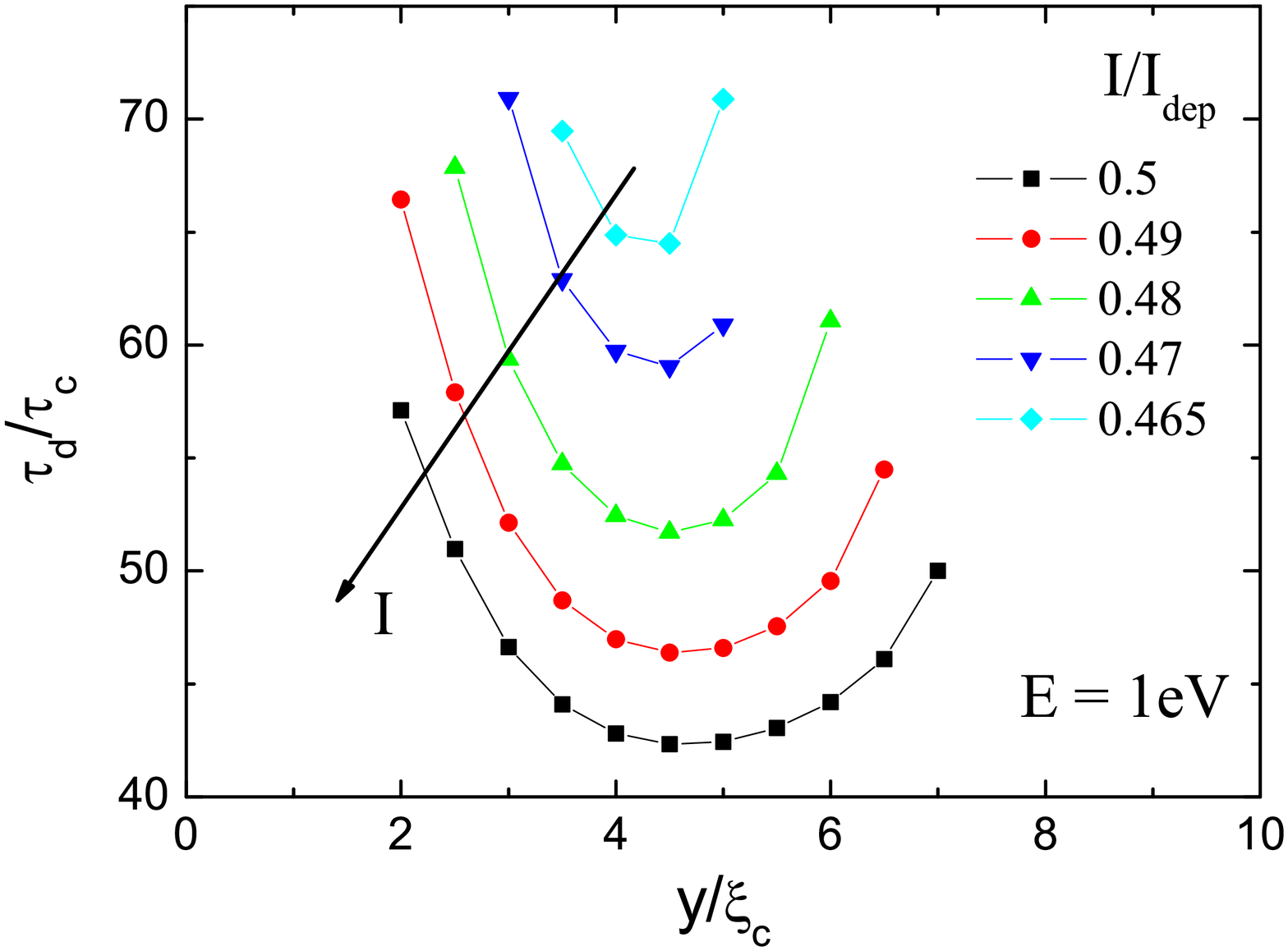}
\caption{Position-dependent delay time at currents close to
$I_{det}^{min} \simeq 0.460 I_{dep}$. We present results only for
left half of the strip, in the right half $\tau_d(y)$ is
symmetric.}
\end{figure}

\section{Discussion}

We do not compare quantitatively our numerical results with
available experiments on dependence of timing
jitter on the current and energy of the photon
\cite{Sidorova,Caloz,Korzh} because we believe that used
theoretical model may give at most semi-quantitatively
correct results. The assumption of the used
model (complete thermalization in electron system) is
fulfilled only partially due to relatively large inelastic
electron-electron relaxation $\tau_{ee}$ for electrons with
energy about $|\Delta|$ above the Fermi level.
As a result the electron distribution function
deviates from Fermi-Dirac distribution with effective
temperature $T_e$ which should affect $\tau_d$.
For example in Ref. \cite{Vodolazov_time_delay} two
limiting cases were considered: complete thermalization
of electrons (quasiequilibrium model in notations of
Ref. \cite{Vodolazov_time_delay}) and nonthermalized
distribution function (nonthermal model). It was found
different (but qualitatively similar) dependencies
$\tau_d(I)$ and $\tau_d$ was shorter in case of thermalized
electrons (compare Figs. 3 and 6 in \cite{Vodolazov_time_delay}).
Therefore we make only semi-quantitative comparison of our
results with available experiments.

In Ref. \cite{Korzh} the monotonous decay of the timing jitter
with current is found for wide range of $E_{\nu}$ and widths of
NbN strips (similar effect is found for MoSi meanders in
\cite{Caloz}). According to our result this effect is connected
with decreasing of the delay time as current increases - effect
comes out from the current dependent relaxation time of $|\Delta|$
\cite{Tinkham_time_delay,Pals,Geier,Jelila,Vodolazov_time_delay}.
Because $\tau_d$ is function of ratio $I/I_{det}$ and $I_{det}$
decreases with increasing of $E_{\nu}$ the delay time and timing
jitter is smaller at fixed current for larger $E_{\nu}$ - this
effect is observed in \cite{Korzh}. Estimation of the depairing
current for 100 nm wide strip from \cite{Korzh} gives us $I_{dep}
\simeq 45 \mu A$ ($T=0.9 K$). It means that the experimental
critical current for this strip ($I_c \simeq 28 \mu A$) is about
of 0.62 $I_{dep}$ and therefore the timing jitter does not reach
its minimal, from theoretical point of view, possible value $\sim
\hbar/k_BT_c \sim 1 ps$ for that NbN strip with $T_c=8.65 K$ (in
Ref. \cite{Korzh} minimal experimental timing jitter is about of
$3 ps$). Sheet resistance for MoSi meanders is not present in Ref.
\cite{Caloz} and we cannot estimate depairing current for studied
structures. Because $T_c$ in MoSi is smaller than in NbN we expect
larger value for minimal timing jitter for this material.

In Refs. \cite{Sidorova,Caloz} nonmonotonous dependence of
jitter on current is observed in the range of currents where
intrinsic detection efficiency (IDE) is smaller than unity.
As we discuss in section V decrease of jitter at relatively
small currents could be connected with decreasing of active
area of detector and/or contribution to photon counts only
absorbed photons with the largest deposited energy $E_{\nu}$.
Does this effect exist in Ref. \cite{Korzh} or not is not clear
because timing jitter is not presented for the currents where IDE $\ll 1$.

The presence of 'shoulder' on dependence of photon counts on $\tau_d$
is a fingerprint of position dependent response.
'Shoulder', qualitatively similar to the one marked in Fig. 6 could
be recognized in supplementary Fig. 8 of Ref. \cite{Korzh},
while in Refs. \cite{Sidorova,Caloz} it looks absent. We have to
stress that the existence of the 'shoulder' depends on probability
of photon absorption across the strip, and hence, on wavelength
of the photon and its polarization. The 'shoulder' is most visible
when photon absorption does not depend on coordinate,
as we assume in our calculations. From another side relatively
strong Fano fluctuations ($\sigma >0.2 E_{\nu} $ for our parameters) may wash
out this feature. But even in this case the position-dependent response
could be revealed in the experiment with external magnetic field,
where it leads to increasing of the width of dependence $P(\tau_d)$,
and, hence, the timing jitter,
while no 'shoulder' is seen - see Fig. 3 in Ref. \cite{Sidorova_arxiv}.

The main {\it qualitative} difference of our results with
theoretical results found in \cite{Kozorezov_arxiv} for the timing
jitter and delay time is the presence of the 'shoulder' on
dependence of photon counts on delay time. This difference is not
surprising because in Ref. \cite{Kozorezov_arxiv}
position-dependent response was not studied. Besides, there are
also two quantitative differences with the model from Ref.
\cite{Kozorezov_arxiv}: i) we do not have coefficient in front of
time derivative $\partial |\Delta|/\partial t$ in TDGL equation
which is proportional to inelastic $\tau_{ep}$ and/or $\tau_{ee}$
- see Eq. (31) in \cite{Kozorezov_arxiv} and ii) in our model
maximal delay time is finite which is connected with finite
life-time of the hot spot. Coefficient in front of $\partial
|\Delta|/\partial t$ appears due to nonequilibrium effect
connected with variation of $|\Delta|$ in time and leads to
relatively long relaxation time of $|\Delta|$
\cite{Tinkham_time_delay,Geier,Vodolazov_time_delay,Watts-Tobin}.
In the form used in Ref. \cite{Kozorezov_arxiv} it is valid at
condition that the delay time is much larger than
$\min\{\tau_{ep},\tau_{ee}\}$. When this is not the case (as in
Ref. \cite{Korzh} at large currents) its usage overestimates the
delay time as it was first discussed in Ref.
\cite{Tinkham_time_delay} (see Fig. 5 there). In the problem with
response on supercritical current pulse already at $I/I_c \gtrsim
1.8$ the delay time practically does not depend on $\tau_{ep}$ as
it could be seen from Figs. 3,6 in \cite{Vodolazov_time_delay}. In
our model this nonequilibrium effect is already included via term
$\partial (E_0\mathcal{E}_s(T_e,|\Delta|)/\partial t$ (see Eq.
(30) in Ref.\cite{Vodolazov_PRA}) which is equivalent to the term
$\sim\partial |\Delta|^2/\partial t$ in Eq. (6) of Ref.
\cite{Vodolazov_time_delay} as $T\to T_c$. Moreover, our model
automatically takes into account that there is no effect of finite
$\tau_{ep}$ on $\tau_d$ in case of strong external driving force
(proportional in this problem to $I/I_{det}$).

The delay time and timing jitter are also calculated in \cite{Wu}
where authors use the model from Ref. \cite{Engel_IEEE_model}.
Disadvantage of this model is connected with the assumption that
vortices enter the strip via the edge of the strip even when the
hot spot is located far from the edge. This assumption comes from
the used in Ref. \cite{Engel_IEEE_model} expression for the energy
barrier for vortex entry which is obtained in framework of the
London model with spatially uniform superconducting order
parameter for straight strip with no hot spot. If one considers
spatial variation of $\Delta$ (using for example Ginzburg-Landau,
Usadel or Eilenberger theories) one immediately finds that the
vortex nucleates in the place where the superconductivity is
maximally suppressed and the supervelocity reaches the maximal
value. For the straight strip with no hot spot the London model
gives correct answer (up to some numerical coefficient) for the
energy barrier because $\Delta$ is suppressed at the edge and the
supervelocity together with the superconducting current density is
maximal there. In the case when the hot spot is located close to
the edge of the strip supervelocity is also maximal at the edge
(while superconducting current density is maximal in another
place) and vortex enters via the edge \cite{Zotova_SUST}. But when
the hot spot is located far from the edge the supervelocity is
maximal inside the hot spot ($\Delta$ is minimal there) and
vortices (vortex/antivortex pair) nucleate inside the hot spot
\cite{Zotova_SUST}. From some point of view the vortex itself is
good illustration of this phenomena. In the center of the vortex
$\Delta=0$, supervelocity diverges and the superconducting current
density is equal to zero. Indirect confirmation of vortex
nucleation inside the hot spot comes also from the recent
experiment \cite{Korneeva_PRA} where single photon detection with
IDE $\sim 1$ is observed in several micron-wide NbN strips which
cannot be explained by vortex penetration via the edge.

\section{Conclusion}

In the framework of two-temperature model combined with modified
time-dependent Ginzburg-Landau equation we find following:

i) delay time and variation of delay time (timing jitter) in SNSPD
connected either with position-dependent response or Fano
fluctuations monotonically decreases with increasing of the
current when $I>I_{det}^{max}$ and timing jitter may be about of
$\hbar/k_BT_c$ at the current close to the depairing current. The
effect is connected with fast decrease of relaxation time of the
superconducting order parameter at large currents. At fixed
current the delay time and timing jitter are smaller for photons
with larger energy due to larger ratio $I/I_{det}$.

ii) Fano fluctuations and nonlinear dependence of $\tau_d(E)$ provide
non-Gaussian dependence of photon counts on delay time,
most pronounced at larger $\tau_d$. Position-dependent response
leads to the appearance of the 'shoulder' on this dependence connected with
contribution from the photons absorbed in near-edge area of the strip.
The 'shoulder' decreases with the current and it is maximal
in case of coordinate-independent photon absorption across the strip.

\begin{acknowledgments}
The work was supported by the Russian Science Foundation (RSF), grant No. 17-72-30036.
\end{acknowledgments}

\end{document}